\documentclass[aps,prl,amsmath,amssymb,twocolumn,superscriptaddress,letterpaper]{revtex4}
\usepackage{dcolumn}
\usepackage{graphicx}
\usepackage{epstopdf}
\usepackage{verbatim}
\usepackage{amssymb}
\usepackage{amsmath}
\usepackage{subfigure,wrapfig}
\usepackage{color}

\begin{document}

\title{Silicon Color Filters for Light-Recycling Displays}

\author{Emma C. Regan$^{1,2}$, Yichen Shen$^{1}$, Aviram Massuda$^{1}$, Owen D. Miller$^{3}$, Marin Solja\v{c}i\'{c}}

\affiliation{
\normalsize{Research Laboratory of Electronics, Massachusetts Institute of Technology,}\\
\normalsize{Cambridge, MA 02139, USA}\\
\normalsize{$^{2}$Applied Science and Technology Graduate Group, University of California Berkeley}\\
\normalsize{Berkeley, CA 94709, USA}\\
\normalsize{$^{3}$Applied Physics and Energy Sciences Institute, Yale University,} \\ 
\normalsize{New Haven, CT 06520, USA}\\
\normalsize{Corresponding author: ycshen@mit.edu} 
}

\begin{abstract}

We propose a set of silicon reflective transmission filters for light-recycling displays that are compatible with high-throughput nanofabrication technologies and have excellent angular stability. 

\end{abstract}

\maketitle

\begin{figure}[t!]
\centering
\includegraphics[width=\linewidth]{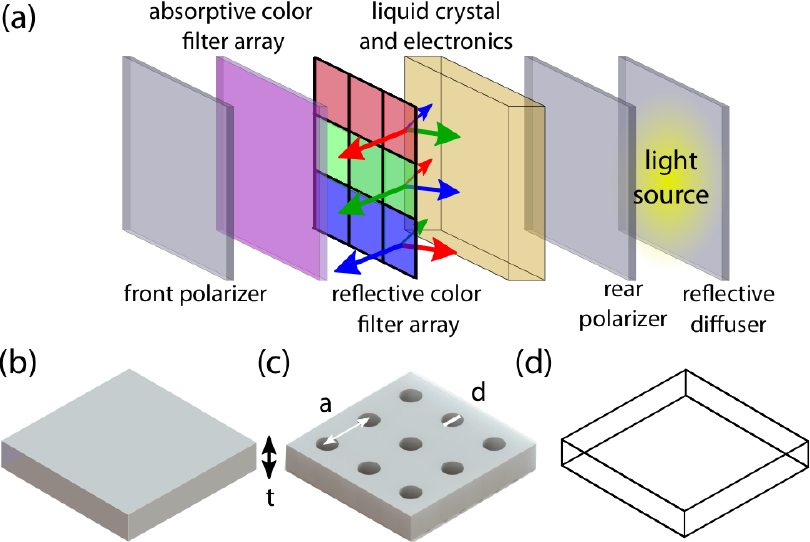}
\caption{(a) Schematic of a liquid crystal display with RT filters. If the RT filters are ideal (transmit 100\% of one color, reflect 100\% of the other two colors), the absorptive filter array is not needed. Each color filter can be (b) a thin layer of c-Si characterized by thickness $t$, (c) a photonic crystal slab with holes of diameter $d$, periodicity $a$, and thickness $t$, or (d) a blank space.}
\label{fig:schematic}
\end{figure}

The backlight in a liquid crystal display [Fig. \ref{fig:schematic}(a)] uses significant power in commonly-used devices. In most situations, a smartphone backlight consumes more power than any other functional component of the phone \cite{Carroll:2010:APC:1855840.1855861}. Beyond personal electronics, industrial displays for advertising and entertainment also consume an enormous amount of energy. To increase display efficiency, a reflective diffuser is typically used to recycle light directed to the back of the display panel by a reflective rear polarizer \cite{Li:09}. However, light recycling has not been fully explored. Common filters based on dyes and pigments absorb more than 70\% of the backlight power instead of recycling it.

While light-recycling transmission filters could save energy, effective and low-cost filters do not yet exist. Planar, multi-layer dielectric stacks (Bragg filters) can serve as efficient reflective transmission (RT) filters and have been used for resonant enhancement in many optical devices \cite{:/content/aip/journal/jap/78/2/10.1063/1.360322}. However, each color requires a different thickness and number of layers, so they are incompatible with high-throughput fabrication methods. Plasmonic filters are an interesting alternative, but they are largely absorptive due to the metal components. \cite{doi:10.1021/ph500410u}. 

In this work, we take advantage of the low absorption and high refractive index of crystalline silicon (c-Si) \cite{VUYE1993166} to design energy-saving RT filters. We optimized a patterned c-Si slab on a quartz substrate, while restricting the thickness to be the same for all filters [Fig. \ref{fig:schematic}(b-d)] \cite{Kanamori:14}, so the filters are compatible with existing high-throughput nanofabrication platforms. The interaction of visible light with the filters provides color selection and light recycling over standard dye and pigmentary filters \cite{doi:10.1021/ph500400w}. 

\begin{figure}[t!]
\centering
\includegraphics[width=\linewidth]{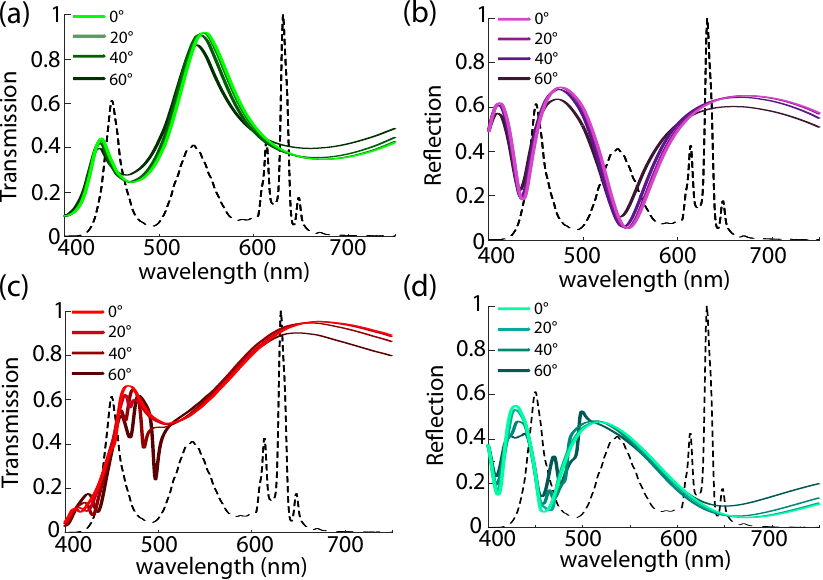}
\caption{Rigorous coupled wave analysis (RCWA) simulations of the transmission and reflection of green filter (a,b) and red filter (c,d) for light of various incident angles showing good angular stability. Incident light is randomly polarized. The black, dashed line in each shows the spectrum of a backlight LED, which can be measured from an iPhone 6s display.}
\label{fig:spectra}
\end{figure}

We computed the transmission and reflection properties of the silicon-on-quartz filters using rigorous coupled wave analysis (RCWA) for layered periodic structures \cite{Liu20122233}. The results indicate that using high-quality c-Si, instead of commonly-studied sputtered metals, significantly enhances the filter quality. A common LED backlight emission spectrum [Fig. \ref{fig:spectra}, dashed lines] has discrete peaks in the red, green, and blue wavelength ranges, which allows for a cleanly defined optimization figure of merit $F$. For example, for a green filter, $F=r(R_B+R_R)-(1-T_G),$ where $R_B$ and $R_R$ are the average reflected blue ($425<\lambda<480$) and red ($600<\lambda<655$) light, and $T_G$ is the average transmitted green ($505<\lambda<575$). $r$ is a measure of the efficiency of the recycling process, and is defined as the fraction of initially reflected light that reaches the color filters a second time. 

We chose the thickness, lattice symmetry, and hole parameters as our degrees of freedom and maximized $F$ using a nonlinear-optimization package NLopt \cite{nlopt}. We first performed a global optimization \cite{1424197} to avoid poor-quality local optima, and then we refined the parameters with a subsequent local optimization algorithm \cite{Powell1994} to achieve a final design. Throughout, we constrained the hole diameter to be larger than $100~\text{nm}$ to achieve realistic and fabricatable photonic crystal slabs.

The optimization produced a set of red and green c-Si color filters with a thickness of $t=134~\text{nm}$ [Fig. \ref{fig:spectra}]. The green filter is a simple slab [Fig. \ref{fig:schematic}(b)]. The red filter is a slab with holes of diameter $d=112~\text{nm}$ and periodicity $a=137~\text{nm}$ in a square lattice. The air holes reduce the effective index of the slab, which shifts the Fabry-Perot resonance to a shorter wavelength. In addition, the hole array supports optical resonances that alter the reflection and transmission spectra via Fano effects \cite{PhysRevB.65.235112,doi:10.1021/ph500400w}. The filters have excellent angular stability due to the high index of silicon, and the Fano resonances in the red filter further improve the filter quality for large angles [Fig. \ref{fig:spectra}]. Regardless of thickness, there was no optimum blue filter due to a spike in the absorption of c-Si near $\lambda=350~\text{nm}$, so the blue pixel is left blank. 

\begin{figure}[t!]
\centering
\includegraphics[width=\linewidth]{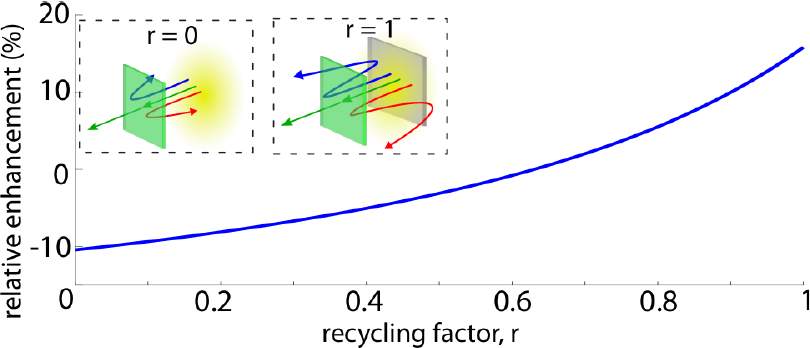}
\caption{The brightness enhancement as a function of the recycling factor $r$ for an RT filter array relative to a standard filter array in a light-recycling display: $\text{relative enhancement} = 100(E_{RT}/E_{\text{standard}}-1)$. In this model, a standard color filter transmits 100\% of one color and absorbs 100\% of the other two colors. For $r>0.6,$ the RT filter increases the brightness.}
\label{fig:performance}
\end{figure}

Including RT filters before the absorptive color filter array increases the brightness of the screen for a given backlight power and therefore reduce the required energy. To quantify this improvement, we consider an LCD panel displaying a white image, so all filters are in the on state. At each wavelength, the enhancement $E$ for an RT color filter can be expressed as a geometric series, $E = (1-R_0)/(1-R_0 r)$, where $R_0$ is the angle-integrated amount of light reflected back for recycling. Considering viewing angles of up to $60$ degrees, the relative brightness enhancement is shown in Figure \ref{fig:performance}. Using a single layer of c-Si filters before the standard absorption filters can enhance the brightness by up to 15\%, depending on the efficiency of the recycling process, which is characterized by the recycling factor, $r$. The recycling factors of current state-of-the-art devices are around $0.5$, but this work suggests that improving $r$ is crucial for reducing display energy consumption.

Furthermore, our design is potentially scalable and compatible with standard high-throughput silicon nanofabrication techniques. We propose the use of commercially available silicon-on-quartz (SOQ) wafers. For each filter set, the green filter can be left unpatterned, the blue filter area can be etched away completely, and the red filter can be patterned with interference \cite{LPOR:LPOR200810061} or immersion lithography \cite{doi:10.1117/12.2021397}.
 
In this letter, we suggest that light recycling via RT color filters can have substantial consequences for energy savings, even with simple silicon designs. 

\section*{Funding Information}
This work was partially supported by the Army Research Office through the Institute for Soldier Nanotechnologies under contract nos. W911NF-13-D0001. MS (reading and analysis) was supported by S3TEC, an Energy Research Frontier Center of the Department of Energy, under grant no. DE-SC0001299.

\bibliographystyle{unsrt}
\bibliography{filter}

\begin{thebibliography}{10}

\bibitem{Carroll:2010:APC:1855840.1855861}
Aaron Carroll and Gernot Heiser.
\newblock An analysis of power consumption in a smartphone.
\newblock In {\em Proceedings of the 2010 USENIX Conference on USENIX Annual
  Technical Conference}, USENIXATC'10, pages 21--21, Berkeley, CA, USA, 2010.
  USENIX Association.

\bibitem{Li:09}
Yan Li, Thomas~X. Wu, and Shin-Tson Wu.
\newblock Design optimization of reflective polarizers for lcd backlight
  recycling.
\newblock {\em J. Display Technol.}, 5(8):335--340, Aug 2009.

\bibitem{:/content/aip/journal/jap/78/2/10.1063/1.360322}
M.~Selim Ünlü and Samuel Strite.
\newblock Resonant cavity enhanced photonic devices.
\newblock {\em Journal of Applied Physics}, 78(2):607--639, 1995.

\bibitem{doi:10.1021/ph500410u}
Zhongyang Li, Serkan Butun, and Koray Aydin.
\newblock Large-area, lithography-free super absorbers and color filters at
  visible frequencies using ultrathin metallic films.
\newblock {\em ACS Photonics}, 2(2):183--188, 2015.

\bibitem{VUYE1993166}
G.~Vuye, S.~Fisson, V.~Nguyen Van, Y.~Wang, J.~Rivory, and F.~Abelès.
\newblock Temperature dependence of the dielectric function of silicon using in
  situ spectroscopic ellipsometry.
\newblock {\em Thin Solid Films}, 233(1):166 -- 170, 1993.

\bibitem{Kanamori:14}
Yoshiaki Kanamori, Toshikazu Ozaki, and Kazuhiro Hane.
\newblock Reflection color filters of the three primary colors with wide
  viewing angles using common-thickness silicon subwavelength gratings.
\newblock {\em Opt. Express}, 22(21):25663--25672, Oct 2014.

\bibitem{doi:10.1021/ph500400w}
Yichen Shen, Veronika Rinnerbauer, Imbert Wang, Veronika Stelmakh, John~D.
  Joannopoulos, and Marin Soljacic.
\newblock Structural colors from fano resonances.
\newblock {\em ACS Photonics}, 2(1):27--32, 2015.

\bibitem{Liu20122233}
Victor Liu and Shanhui Fan.
\newblock S$^4$ : A free electromagnetic solver for layered periodic
  structures.
\newblock {\em Comput. Phys. Commun.}, 183(10):2233 -- 2244, 2012.

\bibitem{nlopt}
Steven~G. Johnson.
\newblock The NLopt nonlinear-optimization package,
  http://ab-initio.mit.edu/nlopt.

\bibitem{1424197}
T.~P. Runarsson and Xin Yao.
\newblock Search biases in constrained evolutionary optimization.
\newblock {\em IEEE Transactions on Systems, Man, and Cybernetics, Part C
  (Applications and Reviews)}, 35(2):233--243, May 2005.

\bibitem{Powell1994}
M.~J.~D. Powell.
\newblock {\em Advances in Optimization and Numerical Analysis}, chapter A
  Direct Search Optimization Method That Models the Objective and Constraint
  Functions by Linear Interpolation, pages 51--67.
\newblock Springer Netherlands, Dordrecht, 1994.

\bibitem{PhysRevB.65.235112}
Shanhui Fan and J.~D. Joannopoulos.
\newblock Analysis of guided resonances in photonic crystal slabs.
\newblock {\em Phys. Rev. B}, 65:235112, Jun 2002.

\bibitem{LPOR:LPOR200810061}
C.~Lu and R.H. Lipson.
\newblock Interference lithography: a powerful tool for fabricating periodic
  structures.
\newblock {\em Laser \& Photonics Reviews}, 4(4):568--580, 2010.

\bibitem{doi:10.1117/12.2021397}
Wim~P. de~Boeij, Remi Pieternella, Igor Bouchoms, Martijn Leenders, Marjan
  Hoofman, Roelof de~Graaf, Haico Kok, Par Broman, Joost Smits, Jan-Jaap Kuit,
  and Matthew McLaren.
\newblock Extending immersion lithography down to 1x nm production nodes.
\newblock volume 8683, 2013.

\end{thebibliography}

\end{document}